\documentclass[nofootinbib,preprint,aps,draft,pre,superscriptaddress,citeautoscript,floatfix]{revtex4-2}
\setcitestyle{super}
\usepackage{graphics}
\usepackage{graphicx}

\usepackage{mathrsfs}
\usepackage{amsmath}
\usepackage{bm}
\usepackage{color}
\usepackage{float}
\usepackage{epstopdf}

\begin{document}

\title{Design and Fabrication of Nano-Particles with Customized Properties using Self-Assembly of Block-Copolymers}
\author{Changhang Huang}
\affiliation{School of Physics, Beihang University, Beijing 100191, China}
\author{Kechun Bai}
\affiliation{School of Physics, Beihang University, Beijing 100191, China}
\author{Yanyan Zhu}
\affiliation{School of Physics, Beihang University, Beijing 100191, China}
\author{David Andelman}
\affiliation{School of Physics and Astronomy, Tel Aviv University, Ramat Aviv, Tel Aviv, 69978, Israel}
\author{Xingkun Man}
\email{manxk@buaa.edu.cn}
\affiliation{School of Physics, Beihang University, Beijing 100191, China}
\affiliation{Peng Huanwu Collaborative Center for Research and Education, Beihang University, Beijing 100191, China}

\begin{abstract}

Functional nanoparticles (NPs) have gained significant attention as a promising application in various fields, including sensor, smart coating, drug delivery, and more. Here, we propose a novel mechanism assisted by machine-learning workflow to accurately predict phase diagram of NPs, which elegantly achieves tunability of shapes and internal structures of NPs using self-assembly of block-copolymers (BCP). Unlike most of previous studies, we obtain onion-like and mesoporous NPs in neutral environment and hamburger-like NPs in selective environment. Such novel phenomenon is obtained only by tailoring the topology of a miktoarm star BCP chain architecture without the need for any further treatment. Moreover, we demonstrate that the BCP chain architecture can be used as a new strategy for tuning the lamellar asymmetry of NPs. We show that the asymmetry between A and B lamellae in striped ellipsoidal and onion-like particles increases as the volume fraction of the A-block increases, beyond the level reached by linear BCPs. In addition, we find an extended region of onion-like structure in the phase diagram of A-selective environment, as well as the emergence of an inverse onion-like structure in the B-selective one. Our findings provide a valuable insight into the design and fabrication of nanoscale materials with customized properties, opening up new possibilities for advanced applications in sensing, materials science, and beyond.

\end{abstract}

\maketitle

\section{Introduction}

Nanoparticles (NPs) have gained significant scientific interest~\cite{Lu01, Saito07, Tanaka09, Shin20, Wong20, Kim24} in recent decades because of their highly promising emerging applications, including sensors~\cite{Choi14}, smart coating~\cite{Shin18}, drug delivery systems~\cite{Venkataraman11}, photonic crystals~\cite{Yang18,Song19}, and more. NPs with abundant shapes and inner structures, such as lamellae (striped ellipsoidal and onion-like)~\cite{Zhu22, Zhu23}, cylinders~\cite{Yang16, Lee19}, perforated lamellae~\cite{Jeon07}, and tulip-bulb-like~\cite{Jeon08, Dai21}, have been produced by utilizing the self-assembly of block copolymers (BCPs). Recently, a series of experiments successfully fabricated striped ellipsoidal NPs with asymmetric lamellae by employing block-selective swelling and regioselective seeded polymerization~\cite{Navarro22-1, Navarro22-2}. Most of current NP fabrications are using linear BCPs and require multi-step processes~\cite{Navarro22-1, Navarro22-2, Kim21, Hu21, Kwon21, Kim22, Shin17}. It is still a challenge to achieve NPs with desired structure and property in a robust fashion.

The topology of the BCP chains can be used to provide a concise self-assembly approach to address this challenge. Modern synthesis techniques enable the precise BCP synthesis with rich architectures and compositions, having the advantage of generating nanomaterials with tunable nanoscale-domain geometry, packing symmetry, and chemical composition~\cite{Bates12, Cook19, Ji21, Seo21-2, Murphy22}. However, existing simulation tools, such as self-consistent field theory (SCFT), Monte Carlo (MC) and molecular dynamics (MD) simulations, struggle to efficiently adapt appropriate BCP chain architecture for desired NPs from a vast number of candidates. This is because the calculations of NPs formed from the self-assembly of BCPs in solvents are not unit-cell calculations, making the numerical investigations time-consuming and skill-intensive~\cite{Bates94, Lee10, Li13, Huang21}. As a result, it is highly desired to improve the efficiency of numerical calculations, especially for BCPs with complex chain architecture.

In recent years, machine learning was gradually developed to investigate self-assembly of BCPs~\cite{Huang21, Tu20, Zhao21, Aoyagi21}. For example, we note the autonomous construction of block-copolymer phase diagrams by theory-assisted active machine learning~\cite{Zhao21}, and the classification of metastable structures of various diblock and triblock copolymers using 3D convolutional neural networks~\cite{Aoyagi21}. Although the current machine-learning methods can reveal the general trend of the phase diagrams, the accurate determination of phase boundaries is still hard to obtain and exploring unknown phase regions still poses a big challenge.  

In this paper, we develope a machine-learning method associated with SCFT to investigate the role played by chain architecture in self-assembly of NPs from $\rm{A}_{1}(A_{2}B)_{n}$ miktoarm star BCPs. This kind of chain architecuture can be utilized to form anomalous micro-structures or stabilize unstable phases of BCP melts~\cite{Li20,Seo21}. We construct the corresponding phase diagram in both neutral and selective solvent environments. Unlike most of previous studies, we obtain onion-like and mesoporous NPs in neutral environment and hamburger-like NPs in selective environment. We emphasize that such novel phenomenon is obtained only by tailoring the topology of the miktoarm star BCPs without the need for any further treatment. We demonstrate that the SCFT-assisted maching-learning method has a major advantage in exploring alternative and more complex architectures within 3D soft confinement. It elegantly achieves tunability of shapes and internal structures of NPs.

\section{Results and Discussion}

{\bf{Machine-Learning Workflow.}} Our system consists of a mixture of an $\rm{A}_{1}(A_{2}B)_{3}$ miktoarm BCP and a homopolymer (C), where the latter acts as a poor solvent for the BCP. We model the chain architecture of the $\rm{A}_{1}(A_{2}B)_{3}$ miktoarm by using two parameters: the volume fraction of the A component, $f$, and the ratio between the $\rm{A}_{1}$ block and the entire A component, $\tau = f_{A_{1}}/f$. 

\begin{figure*}
{\includegraphics[width=1.0\textwidth,draft=false]{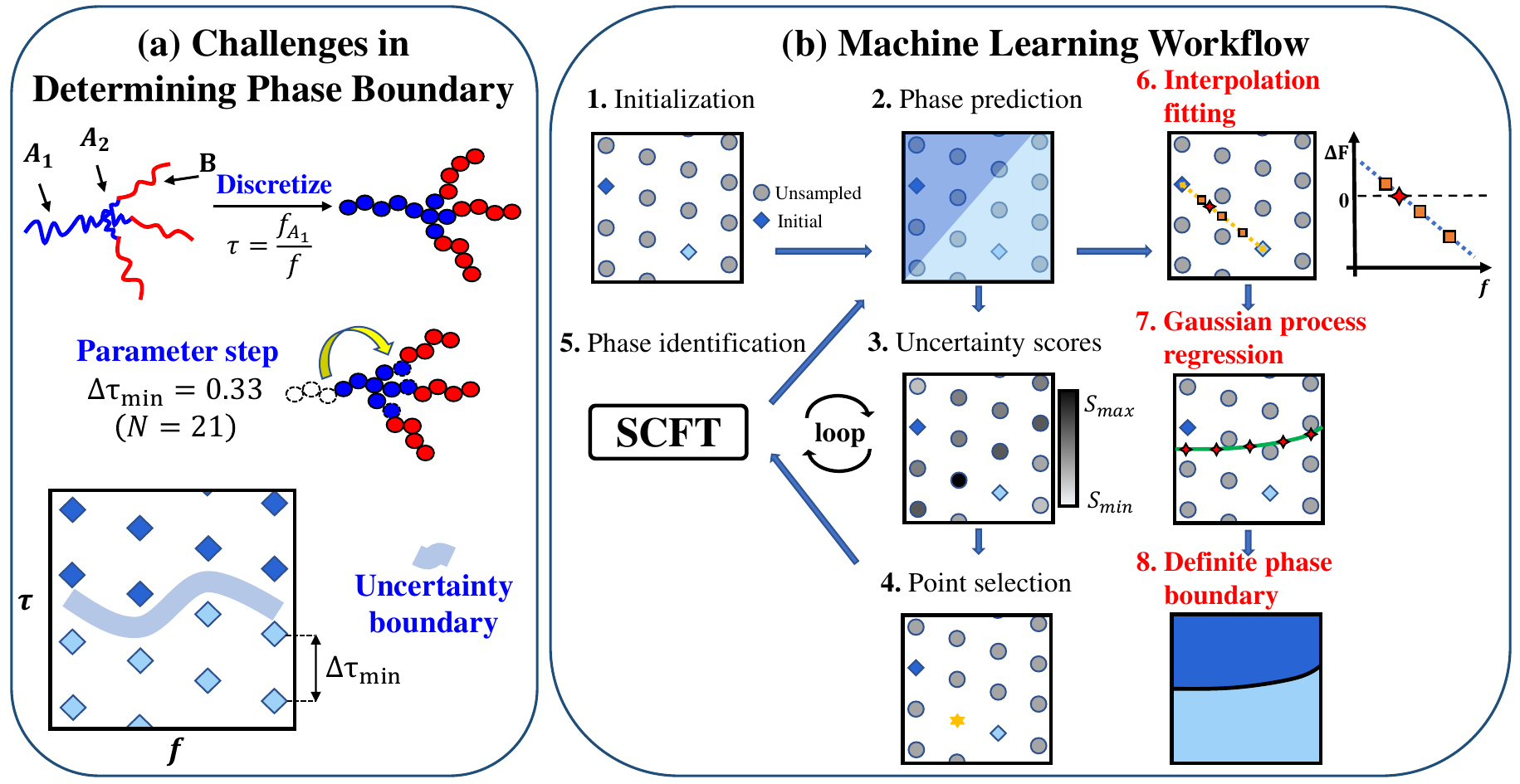}}
\caption{
\textsf{(a) Challenges in determining phase boundaries by conventional methods. Due to the constraint of discretization and chain architecture, the value of parameter $\tau=f_{A_1}/f$ cannot be varied at will. For $N=21$, the minimal step of $\tau$ is $\bigtriangleup\tau=0.33$, leaving a wide uncertainty region (denoted as bluish grey band) between two different phases. (b) Schematic of the machine-learning workflow. Steps 1 to 5 constitute SCFT-assisted active-machine learning, which samples data near the phase boundaries. Subsequently, step 6 employs the $k$-nearest-neighbor algorithm to select points that cross these boundaries (interpolation fitting). Finally, step 7 performs a Gaussian process to obtain the nonlinear regression curve as the definite phase boundary.}}
\label{fig1}
\end{figure*}

To figure out the effect of chain architecture on the BCP particle morphology, we first construct a phase diagram in terms of $f$ and $\tau$ in a {\it neutral environment} where the C homopolymer has no preference towards either A or B component. Note that due to the discretization procedure, the contour steps of $\rm{A}_{2}$ and B-blocks have to be a multiple of three, as the three $\rm{A}_{2}B$ arms have the same length. As a result, traditional numerical methods, such as SCFT, could only be carried out at discrete values of ($\tau$, $f$) with large discrete steps, $\bigtriangleup\tau$ and $\bigtriangleup f$. This implies large unexplored regions in the ($\tau$, $f$) phase diagram. For example, for the degree of polymerization $N=21$ and volume fraction $f=0.43$, the minimal discrete step of $\tau$ is rather large, $\bigtriangleup\tau=0.33$, as shown in Fig.~\ref{fig1}(a). This leads to two difficulties in the construction of the phase-diagram. The first is that the gap between two neighboring points belonging to two different phases is large, making it difficult to determine accurately the boundary between those phases. The second is that phases residing in small regions in the phase diagram may not be detected. Although the numerical accuracy of the phase boundary can be increased by increasing $N$, it causes a considerable increase in computational cost because non-unit cell calculations are required in determining the equilibrium structure of NPs, making it hard to accurately obtain the phase boundaries.

To address this problem, we present a novel machine-learning workflow, which combines SCFT-assisted active-learning loops~\cite{Zhao21}, $k$-nearest-neighbor algorithm~\cite{Cover67}, Gaussian process regression~\cite{Seeger04} and free-energy changes calculated near the phase boundaries. Our machine-learning workflow is shown in Fig~\ref{fig1}(b). First, we employ SCFT-assisted active-machine learning to sample data points near the phase boundaries (steps 1-5 in Fig.~\ref{fig1}(b)). Subsequently, the $k$-nearest-neighbor algorithm is adopted to select points that cross the phase boundaries (step 6). Finally, we utilize a Gaussian process to obtain the nonlinear regression curve as our best estimate for the phase boundary (steps 7 and 8). The performance of this machine-learning procedure has been tested by reproducing the well-studied phase diagram of di-BCP melts, acting as a preliminary test of our method. The accuracy of the calculated phase diagram determined from the $F$-score increases from $F=0.96$ to $0.993$, where the $F$-score is defined as the arithmetic mean of the precision and recall~\cite{Yang99}. More details about the accuracy improvement in constructing such phase diagrams can be found in the Supporting Information (Fig. S1).

\begin{figure*}
{\includegraphics[width=1.0\textwidth,draft=false]{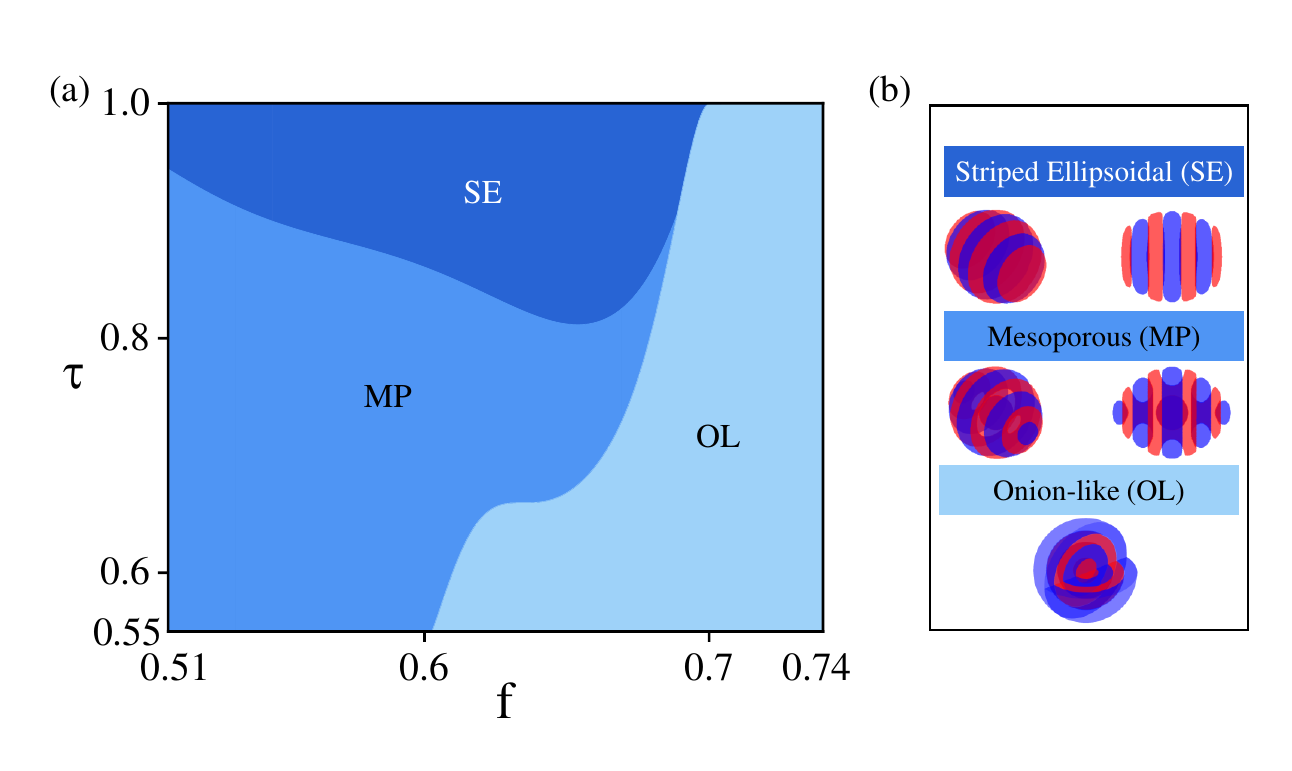}}
\caption{
\textsf{(a) The predicted phase diagram of miktoarm BCP nanoparticles with respect to the chain architecture parameters $\tau$ and $f$. (b) The 3D patterns of the three phases: Striped Ellipsoidal (SE), Mesoporous (MP), and Onion-like (OL). Other parameters are $L_{x}{=}L_{y}{=}L_{z}{=}15$, $N_{x}{=}N_{y}{=}N_{z}{=}64$, $N_s{=}200$, $f_{\rm C}{=}0.3$, $\phi_{0}{=}0.2$, $N\xi {=}300$, $N\chi_{\rm AB}{=}35$, $N\chi_{\rm AC}{=}N\chi_{\rm BC}{=}14$.}}
\label{fig2}
\end{figure*}

{\bf{Self-Assembly of $\rm{A}_{1}(A_{2}B)_{3}$ NPs in Neutral Environment.}} We empoly our machine-learning technique to construct the phase diagram of BCP particles formed by an $\rm{A}_{1}(A_{2}B)_{3}$ miktoarm in {\it neutral environment}, as shown in Fig.~\ref{fig2}. We find that mesoporous (MP) and onion-like (OL) particles can be formed in neutral environment by tunning the chain architecture. For linear BCPs studied previously, the formation of particles with mesoporous or onion-like structure usually require specific conditions: the former requires a selective environment~\cite{Lee19, Zhu22}, while the latter requires extreme asymmetric A/B volume fraction~\cite{Wong20, Lin17}. However, for $\rm{A}_{1}(A_{2}B)_{3}$ miktoarm NPs, these two structures can be formed in neutral environment by adjusting the chain architecture. The neutral environment is obtained by equating the Flory-Huggins parameter between the A and B components of the BCP and the homopolymer (C), $N\chi_{\rm AC} = N\chi_{\rm BC}$. For MP structures, the `A' domains form doughnuts and cores, while the `B' domains coalesce through the holes. For the range of $0.51\le f\le 0.69$, most of the phase diagram is occupied by the MP phase, while the OL phase exists only in the range $0.6\le f\le 0.64$ and small values of $\tau$. However, for larger $f$, the OL phase is always the most stable. With large values of $\tau$ and $0.51\le f\le 0.7$, $\rm{A}_{1}(A_{2}B)_{3}$ miktoarm tend to form conventional striped ellipsoidal (SE) particles, similar to linear di-BCP systems in neutral environment. Our machine-learning method predicts that the boundary between the SE and MP phases decreases from $\tau=0.95$ to 0.81, and then increasing back to 0.91. The boundary between MP and OL increases from $\tau=0.55$ to a plateau at $\tau=0.66$, and then continues to increase to $\tau=1$. These results are clearly seen in Fig.~\ref{fig2}.

\begin{figure*}
{\includegraphics[width=1.0\textwidth,draft=false]{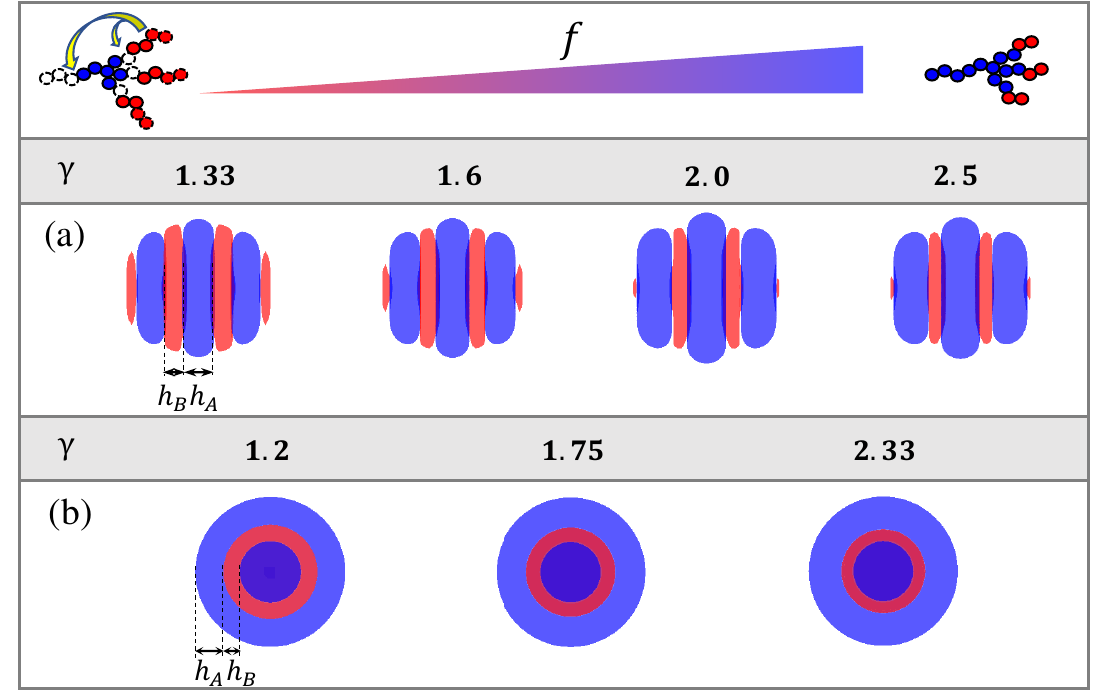}}
\caption{
\textsf{(a) The lamellar asymmetry in striped ellipsoidal nanoparticles for $\tau{\approx}0.89$ and for an increasing series of $f{=}0.565, 0.61, 0.655, 0.685$. The value of the corresponding asymmetry parameter $\gamma{=}h_{A}/h_{B}$ is listed above the particles. (b) The asymmetry in onion-like nanoparticles for ${\tau}{\approx}{0.85}$ and $f{=}0.685, 0.715, 0.745$ (with corresponding $\gamma$ values). Other parameters are $L_{x}{=}L_{y}{=}L_{z}{=}15$, $N_{x}{=}N_{y}{=}N_{z}{=}64$, $N_s{=}200$, $f_{\rm C}{=}0.3$, $\phi_{0}{=}0.2$, $N\xi {=}300$, $N\chi_{\rm AB}{=}35$, $N\chi_{\rm AC}{=}N\chi_{\rm BC}{=}14$.}}
\label{fig3}
\end{figure*}

The lamellae exhibit high asymmetry of A/B domain thickness for the SE and OL NPs. We find that the lamellar asymmetry increases with increasing $f$. Figure~\ref{fig3} shows the dependence of lamellar asymmetry in SE ($\tau\approx0.89$) and OL ($\tau\approx0.85$) nanoparticles on the chain architecture parameter $f$. The lamellar asymmetry is characterized by the thickness ratio of A and B domains, $\gamma=h_{\rm{A}}/h_{\rm{B}}$. This also can be seen in Fig.~\ref{fig3} (a), where a series of SE particles with $\tau\approx0.89$ are shown. As $f$ increases, $f=0.565, 0.61, 0.655,$ and $0.685$, the resulting $\gamma$ also increases, $\gamma=1.33, 1.6, 2.0,$ and $2.5$. We also illustrate a series of OL patterns for $\tau\approx0.85$, and increasing $f=0.685, 0.715,$ and $0.745$, as shown in Fig.~\ref{fig3} (b). It results in the corresponding value of $\gamma$ to increase as well $\gamma=1.2, 1.75,$ and $2.33$. With increasing $f$, the length of the $\rm{B}$-block decreases while the length of $\rm{A}_{1}$ and $\rm{A}_{2}$ blocks increases. Note that the lamellae inside the particles exhibit a gradually increasing asymmetry with increasing $f$, and finally the NPs are composed of quite high asymmetric lamellae. Note that the $\rm{A}_{1}(A_{2}B)_{3}$ miktoarm NPs can form lamellae even at $\gamma=2.5$ and $f=0.685$. This value of $f$ largely exceeds any value where conventional di-BCPs can form lamellar structure~\cite{Huang21, Matsen12}. Hence, our strategy provides a novel way to tailor inner NP structure by designing BCP chain architecture.

Except for accommodating lamellar asymmetry, our calculations indicate that the ellipticity of SE NPs and the number of layers in OL NPs can be tunned by chain architecture. With increasing $\tau$, the ellipticity increases for SE NPs, while for OL NPs, the number of layers decreases. The detailed results and mechanisms are discussed in the Supporting Information (Fig.~S2).

\begin{figure*}
{\includegraphics[width=1.0\textwidth,draft=false]{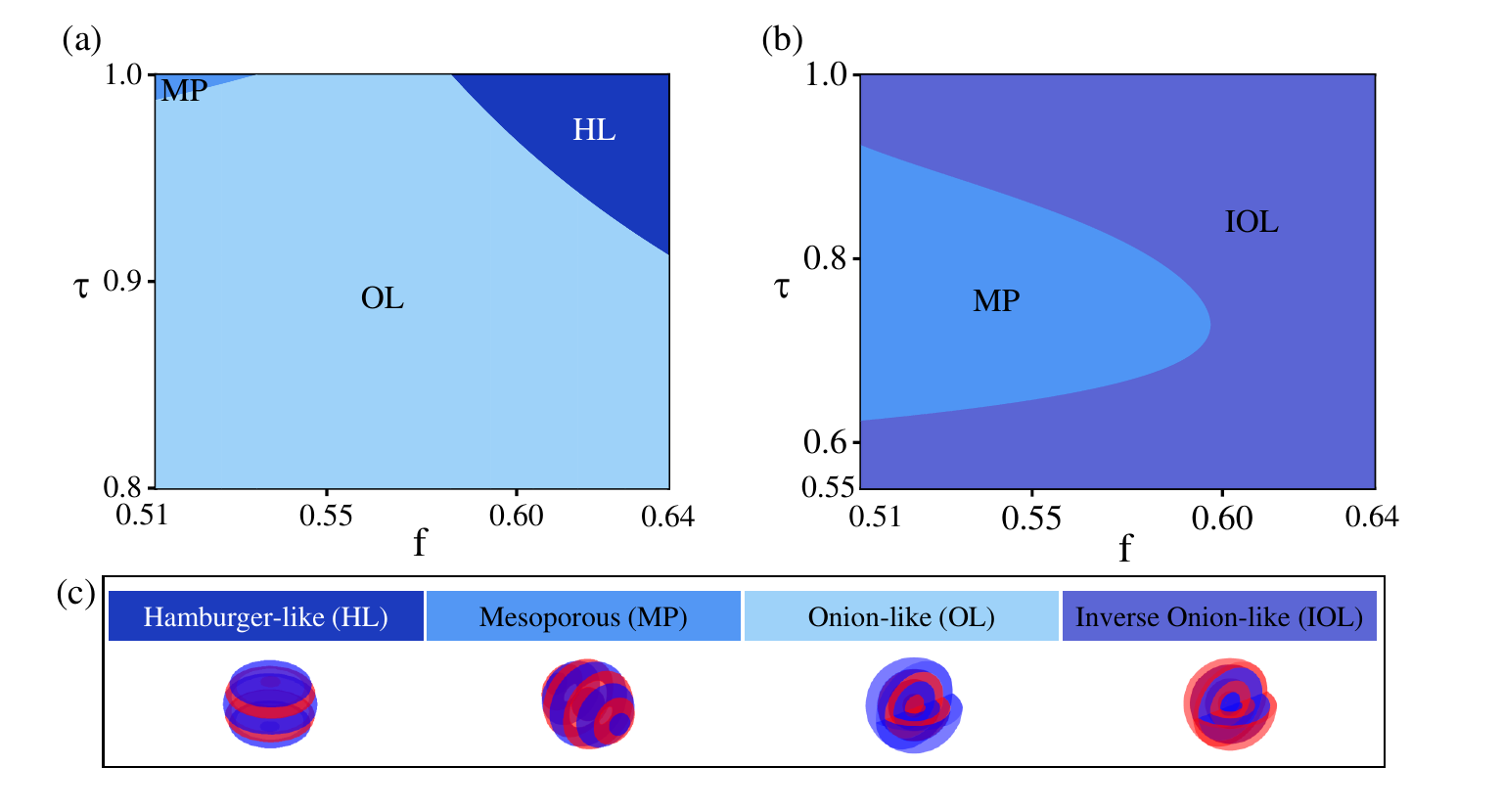}}
\caption{
\textsf{The phase diagram of miktoarm BCP particles in {\it selective environment} plotted in the parameter space of the chain architecture parameters, $\tau$ and $f$. (a) The A component is preferred by the environment, $N\chi_{\rm AC}{=}14 < N\chi_{\rm BC}{=}29$. (b) The B component is preferred by the environment, $N\chi_{\rm AC}{=}29$ $>$ $N\chi_{\rm BC}{=}14$. Other parameters are $L_{x}{=}L_{y}{=}L_{z}{=}15$, $N_{x}{=}N_{y}{=}N_{z}{=}64$, $N_s{=}200$, $f_{\rm C}{=}0.3$, $\phi_{0}{=}0.2$, $N\xi {=}300$, $N\chi_{\rm AB}{=}35$. (c) The 3D patterns of the four phases considered in (a) and (b) are indicated as Hamburger-like (HL), Mesoporous (MP), Onion-like (OL), and Inverse Onion-like (IOL).}}
\label{fig4}
\end{figure*}

{\bf{BCP Nanoparticles in Selective Environment.}} In past experiments, the selectivity of the environment, {\it{i.e.}}, the preference of the aqueous phase towards one of the two (A/B) components was often ultilized to tailor the NP structure~\cite{Zhu22, Lee19, Kim21, Kwon21, Kim22}. In Fig.~\ref{fig4} we show the phase diagram with respect to the chain architecture parameters, $\tau$ and $f$, in A- and in B-selective environments. The corresponding 3D patterns are indicated as well. Herein, the preference is achieved by setting different values of Flory-Huggins parameter between the two BCP components (A and B) and homopolymer (C), $\chi_{\rm AC}\neq\chi_{\rm BC}$.

Figure~\ref{fig4}(a) shows the phase diagram of an A-selective environment ($N\chi_{\rm AC}{=}14 < N\chi_{\rm BC}{=}29$). The hamburger-like (HL) structure, which was observed in previous experiments only for ABC triblock copolymers~\cite{Zhou15, Lu20}, emerges in our studies. When the $\tau$ values approach unity, the MP phase is the most stable phase for $0.5\le f\le 0.54$, while the HL phase is the most stable one for $0.58\le f\le 0.64$. This leads to a phase transition from MP to OL, followed by a HL phase for further increased $f$. Compared with the neutral environment in Fig.~\ref{fig2}, the phase diagram of the A-selective environment shows that most of the phase diagram is occupied by OL structure, which has a lower surface energy. However, although the region occupied by the MP phase is very small, it is found by our machine-learning method.

Figure~\ref{fig4}(b) shows the phase diagram of a B-selective environment, with $N\chi_{\rm AC}{=}29 > N\chi_{\rm BC}{=}14$. Unlike Fig.~\ref{fig4}(a), here an inverse onion-like (IOL) structure emerges, where the outermost layer is composed of the B domain. For $0.51\le f\le 0.6$, MP phase is mostly stable for moderate values of $\tau$, while larger or smaller $\tau$ value lead to a phase transition from a MP phase to an OL one. Compared with neutral and A-selective environments, MP structure still occupied a relatively large area in the phase diagram. However, the SE or HL structures completely disappear in the B-selective environment. The asymmetry characteristic to the two phase diagrams of Fig.~\ref{fig4} for A/B selective environment is attributed to the asymmetric arrangement of A/B components in the chain architecture of $\rm{A}_{1}(A_{2}B)_{3}$ miktoarms.

{\bf{Mechanisms of Chain-Architecture-Induced Structural Transition.}} We find that $\rm{A}_{1}(A_{2}B)_{3}$ miktoarms can form OL NPs in neutral environment by solely modifying the chain architecture. Generally, OL NPs are only observed in selective environments, while SE NPs with flat A/B interface are obtained in a neutral environment. Such counter-intuitive result is attributed to the spontaneous curvature of A/B interface associated with an inherent molecular architecture of the miktoarm. This can be understood in the following way. In the limit $\tau \rightarrow 1.0$, the $\rm{A}_{1}(A_{2}B)_{3}$ miktoarm reduces to an $\rm{AB}_{3}$ star BCP that has no spontaneous curvature. However, the emergence of $A_{2}$ block in the three arms for $\tau<1$ can generate spontaneous curvature of the A/B interface~\cite{Seo21}. Moreover, the spontaneous curvature becomes larger for smaller values of $\tau$ because the length of $A_2$ increases as $\tau$ decreases, leading to a phase transition from the preferred SE (flat A/B interface) to MP, and ultimately to OL NP (curved A/B interface). The effect of $f$ on NP is the same as that of $\tau$. As the value of $f$ deviates away from $0.5$, the spontaneous curvature of A/B interface increases~\cite{Shi21}, leading to a phase transition from SE to OL NP as shown in Fig.~\ref{fig2}. Our finding indicate that the phase diagram reflects the competition between the spontaneous curvature and the neutral environment conditions. The former prefers OL particle, while the latter prefers SE particle. 

\begin{figure*}
{\includegraphics[width=1.0\textwidth,draft=false]{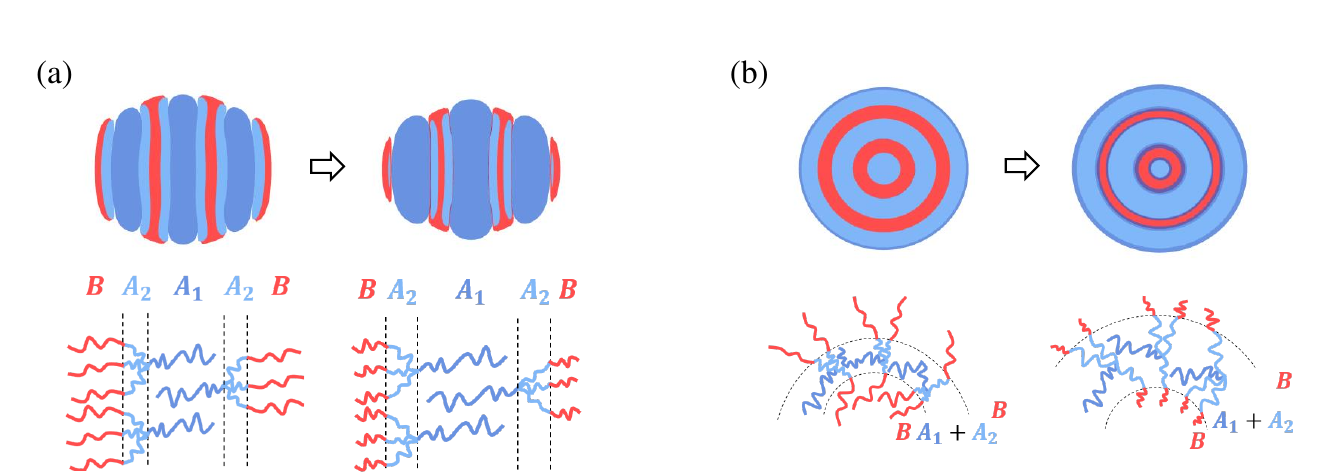}}
\caption{
\textsf{(a) The density distribution in the SE phase of $\rm{A}_2$-blocks for $f{=}0.505$, $\tau{=}0.97$ (left); and $f{=}0.64$, $\tau{=}0.97$ (right), with the corresponding schematic arrangement of BCP chain inside the particles. (b) The density distribution in the OL phase of $\rm{A}_2$ blocks for $f{=}0.61$, $\tau{=}0.5$ (left); and $f{=}0.64$, $\tau{=}0.5$ (right). The corresponding arrangement of BCP chains inside the particles is shown schematically. Other parameters are $L_{x}{=}L_{y}{=}L_{z}{=}15$, $N_{x}{=}N_{y}{=}N_{z}{=}64$, $N_s{=}200$, $f_{\rm C}{=}0.3$, $\phi_{0}{=}0.2$, $N\xi {=}300$, $N\chi_{\rm AB}{=}35$, $N\chi_{\rm AC}{=}N\chi_{\rm BC}{=}14$.}}
\label{fig5}
\end{figure*}

The emergence of asymmetric lamellae in SE NPs is attributed to an {\it{effectvie length}} of the A component, $l_{\rm{eff}}$. To reveal the chain arrangement inside the NPs with asymmetric lamellae, we calculate the density distribution of the $\rm{A}_2$ blocks for SE particles in two cases ($f{=}0.505$, $\tau{=}0.97$ and $f{=}0.64$, $\tau{=}0.97$), as shown in Fig.~\ref{fig5}(a). The $\rm{A}_2$ blocks (light blue) concentrate in the A-domain edge, while the domain center is mainly occupied by the $\rm{A}_1$ blocks (blue). This means that three $\rm{A}_{2}B$ miktoarm arms are arranged along the same direction, as shown schematically in the figure. Hence, we can define an effectvie length of A component, $l_{\rm{eff}}=N_{A1}+N_{A2}/3$. When $f$ increases, the symmetry of A and B components is broken. However, as the four A segments share the increase of the A component length, the increase of the A-component effectvie length is not significant. As a result, the particle maintains its lamellar structure, but the asymmetry of the characteristic length of $\rm{A}$ and $\rm{B}$ domains increases.

For OL nanoparticles, the asymmetric lamellae is attributed to the effect of a {\it{bridge length}} $l_{\rm{br}}$. The density distribution of $\rm{A}_2$ blocks for $f=0.61$, $\tau=0.5$ and for $f=0.64$, $\tau=0.5$ show that $\rm{A}_2$ blocks distribute uniformly in the A domains, as presented in Fig.~\ref{fig5}(b). This means that the $\rm{B}$-block in the three arms of one miktoarm BCP is arranged into two neighboring $\rm{B}$ domains. We then assert that the $\rm{A}_2$ segments act as a bridge connecting the $\rm{B}$ domains, and the width of the $\rm{A}$ domains ({\it{i.e.}}, the distance between two neighboring $\rm{B}$ domains) depends mainly on the bridge length, $l_{\rm{br}}=2N_{A2}$. With increasing $f$, the bridge length $l_{\rm{br}}$ increases, resulting in the increase of characteristic length of $\rm{A}$ domain that facilitates the formation of asymmetric lamellae in OL particles.

\section{Conclusions}

We investigate the effect of chain architecture on the self-assembly and morphology of nanoparticles using $\rm{A}_{1}(A_{2}B)_{3}$ miktoarm star copolymers, and propose a novel machine-learning workflow. A phase diagram is constructed for {\it neutral} as well as {\it selective} environments with respect to the two A/B copolymer components. The neutral environment phase-diagram reveals three distinct structures: striped ellipsoidal (SE), mesoporous (MP), and onion-like (OL) particles. Interestingly, unlike previous findings on linear BCP chains, our calculations demonstrate that by modifying the BCPs' chain architecture, MP and OL structures can be stabilized in a neutral environment with moderate volume fractions. By decreasing $\tau$ and increasing $f$, the morphology of BCP particles changes from SE to MP and finally to OL particles. 

Furthermore, employing SCFT, we demonstrate that it is possible to tune the asymmetry of lamellae in SE and OL particles by modifying the chain architecture of the $\rm{A}_{1}(A_{2}B)_{3}$ miktoarm star copolymer. (i) For SE particles and $0.82\le \tau \le 1$, the lamellar asymmetry can be enhanced by increasing $f$, where the insufficient increase of the A component {\it{effectvie length}} (defined above) maintains the lamellar structure. (ii) For OL particles and $0.52 \le \tau \le 1$, the lamellar asymmetry can be enhanced by increasing $f$, where the {\it{bridge length}} (above introduced) broaden the $\rm{A}$ domain width. Moreover, the ellipticity of SE particles and the number of layer of OL structures can be tuned by $\tau$. SE particles become more prolate with increasing $\tau$, while for OL particles, the number of layer decreases with increasing $\tau$. 

We also construct phase diagrams for A-selective and B-selective environments in order to take into account the environment selectivity. In the A-selective environment, the OL phase region is significantly expanded and a hamburger-like (HL) phase emerges. In the B-selective environment, in contrast, we observe the formation of inverse onion-like (IOL) structures and disappearance of the SE structures.

Our study brings two major advantages: (i) a novel machine-learning workflow to accurately predict phase diagram in polymer science; (ii) highlighting the importance of chain architecture in controlling morphologies and inner structures of block copolymer particles, and shedding light on the principles behind self-assembly of complex nanoscale materials.

\section{Methods}

The calculations are based on self-consistent field theory (SCFT)~\cite{Huang21} and machine-learning tools. For the SCFT calculations, the following parameters are fixed: the size of simulation box $L_{x}\times L_{y}\times L_{z}=15\times 15\times 15$ (in units of $R_g$), discretized into $N_{x}\times N_{y}\times N_{z}=64\times 64\times 64$ lattice sites; the average BCP volume fraction $\phi_{0}=0.2$; the ratio of chain length between the homopolymer and BCP $f_C=0.3$; the number of contour steps along the chain $N_s=200$; the Helfand coefficient, $N\xi =300$; the Flory-Huggins parameter between $\rm{A}$ and $\rm{B}$, $N\chi_{\rm{AB}} = 35$ ensuring strong segregation of the A/B domains inside the BCP particle. 

The machine-learning workflow consists of SCFT assisted active-learning loops~\cite{Zhao21}, $k$-nearest neighbor algorithm~\cite{Cover67}, and Gaussian process regression~\cite{Seeger04} on the basis of the free-energy difference near the phase boundaries. The sampling points are selected on a $10\times 15$ grid after active-learning cycles, and the final phase diagram is predicted on a $801\times 801$ grid. The number of neighbors selected by the $k$-nearest neighbor algorithm is $k=3$. 

More detail in SCFT and machine-learning procedures can be found in Supporting Information.

\bigskip
{\bf Acknowledgement.}~~
We thank Y. Jiang and Q. Dong for useful discussions. This work was supported in part by the NSFC-ISF Research Program, jointly funded by the National Natural Science Foundation of China (NSFC) under grant no. 21961142020 and the Israel Science Foundation (ISF) under grant no. 3396/19, NSFC grant no. 21822302, and ISF grant no. 213/19, and the Fundamental Research Funds for the Central University under grant no. YWF-22-K-101, and the ``111 Center" (No. 20065). The authors also acknowledge the support of the High-Performance Computing Center of Beihang University.

\newpage



\end{document}